\pdfoutput=1 
\documentclass[12pt,preprint]{aastex}
\usepackage{natbib}

\def \arcsec {\hbox{$^{\prime\prime}$}}

\begin{document}

\title{Multi-wavelength high-resolution observations of a small-scale emerging magnetic flux event and the chromospheric and coronal response}

\author{Santiago Vargas Dom\'inguez$^{1}$, Alexander Kosovichev$^{1,2,3}$ \& Vasyl Yurchyshyn$^{1}$}
\affil{$^{1}$ Big Bear Solar Observatory, NJIT, 40386 North Shore Lane, Big Bear City, CA 92314-9672, U.S.A. \\ 
              $^{2}$ Stanford Unversity, Stanford, CA 94305, U.S.A.\\
              $^{3}$ NASA Ames Research Center, Moffett Field, CA 94035, U.S.A.\\
                }
 

\begin{abstract}
State-of-the-art solar instrumentation is now revealing magnetic activity of
the Sun with unprecedented temporal and spatial resolutions. Observations with the 1.6m aperture New Solar Telescope
(NST) of the Big Bear Solar Observatory (BBSO) are making next steps in our
understanding of the solar surface structure. Granular-scale
magnetic flux emergence and the response of the solar atmosphere are among the key research topics of high-resolution solar physics. As part of a joint observing program with NASA's Interface Region Imaging Spectrograph  (IRIS) mission, on August 7, 2013,
the NST observed active region NOAA 11810 in the photospheric TiO 7057\AA~band with a resolution of 0.034\arcsec/pix, and chromospheric He\textsc{i} 10830\AA~
and H$\alpha$ 6563\AA~wavelengths. Complimentary data are provided by Solar Dynamics Observatory (SDO) and Hinode space-based telescopes. The region displayed a group of solar pores,
in the vicinity of which we detect a small-scale buoyant horizontal
magnetic flux tube causing abnormal granulation and interacting with the pre-existing ambient field in upper atmospheric layers.
Following the expansion of distorted granules at the the emergence site, we observed a sudden appearance of an extended surge in the HeI 10830\AA~data
(bandpass of 0.05\AA). The IRIS transition region imaging catched ejection of a hot plasma jet associated with the HeI-surge. The SDO/HMI data used to study the evolution
of the magnetic and Doppler velocity fields reveal emerging magnetic looplike structures. \emph{Hinode}/Ca\textsc{ii}~H and IRIS 
filtergrams detail the connectivities of the newly emerged magnetic field in the lower solar chromosphere. From these data we find that the orientation
of the emerging flux tube was almost perpendicular to the overlying ambient field.  Nevertheless the interaction of emerging magnetic field lines with the pre-existing overlying field 
generates high-temperature emission regions and boosts the surge/jet production. The localized heating is detected before and after the first signs of the surge/jet ejection. We compare the results with previous observations and theoretical models, and propose a scenario for the activation of plasma jet/surges and confined heating triggered by buoyant magnetic flux tubes rising up into a magnetized upper environment. Such process may play a significant role in the mass and energy flow from the interior to the corona.
\end{abstract}

\section{Introduction}
Activity of the solar atmosphere entails numerous multi-scale processes, magnetic structuring of which is controlled by photospheric and subphotospheric evolution and dynamics. Building blocks for these processes are thought to occur at very small (subarsecond) spatial and short (few minutes) temporal scales. The 1.6 m New Solar Telescope \cite[NST,][]{goode2010} operating at the Big Bear Solar Observatory provides such high-resolution capabilities. The ground-based observations reaching the diffraction limit through the use of adaptive optics systems in combination with the current satellite facilities, i.e., IRIS, SDO and Hinode, allow us to investigate the linkage between different layers of the solar atmosphere from the photospheric surface to the corona, in order to detail the finest evolutionary stages of solar activity and to understand the physical mechanisms driving it. Emerging flux regions (EFRs) are of great interest because of their impact on the solar atmosphere. Many observational and theoretical approaches \citep[e.g.][]{zwaan1985,lites1998,magara2001,kubo2003,stein2011} have been developed to establish how magnetic fields are generated in the solar interior, emerge in the photosphere, and shape the structure and dynamics of atmospheric layers. It is thought that magnetic reconnection may play a particularly important role in shaping the response of the solar chromosphere to the emerging magnetic flux associated with different phenomena, such as brighenings, dimmings, jets, surges, among others. Some recent models suggest that flux emergence and subsequent reconnection with the background magnetic field can be important for injection of mass and energy into upper atmospheric layers \citep{archontis2005,isobe2005,martinez2008}. Heating and eruption of chromospheric material associated with granular-scale EFR in vicinity of active regions have been observed in locations displaying surges \citep{guglielmino2010,wang2014}, jets \citep{guo2013,schmieder2013}, Ellerman bombs and various brightenings \citep{pariat2007,vargas2012}. These observations suggested that magnetic reconnection between newly emerged and pre-existing fields can release energy and drive ejection of chromospheric plasma, supporting a simple 2D configuration scenario proposed by some authors \citep[e.g.][]{yokoyama1995}. In this scenario, the emerging magnetic field lines are anti-parallel to the pre-existing field lines. However, in the real 3D geometry the orientation of interacting field lines may be more complex and significantly affects the dynamics and energetics of the process, as pointed out by \cite{galsgaard2007}. The complex dynamics of small-scale and ``hidden'' fields in the photosphere is claimed to be important for balancing radiative energy losses of the chromosphere \citep[e.g.][]{trujillo2004}. However, the key ingredients of the small-scale emerging flux and its interaction with the solar atmosphere are still not understood. In this paper, we present a multi-wavelength analysis of observations, acquired with the NST, the Solar Dynamics Observatory \citep[SDO;][]{lemen2012}, Interface Region Imaging Spectrograph \citep[IRIS;][]{depontieu2014} and the Hinode mission \citep{kosugi2007}, of a transient emerging flux event of 2013 August 7, resulting in generation of a surge/jet and compact heating at coronal-height locations. We compare our observational findings with realistic MHD numerical modeling of magnetic field emergence from the convection zone into the chromosphere and corona. In particular, we find remarkable similarities with the evolution of buoyant flux tubes in three-dimensional numerical experiments of \cite{cheung2007} and \cite{tortosa2009}, in which the equations of MHD and radiative transfer are solved self-consistently. We present an interpretation of our observational  data in terms of the emergence of an horizontal small-scale magnetic flux tube into an overlying magnetized environment, and discuss of a plasma jet generated by this emergence.

\section{Observations and Data Analysis}
The emerging magnetic flux event was observed in the vicinity of active region NOAA AR 11810 at solar disk location S26W18 (263\arcsec,-501\arcsec), on August 17, 2013. The NST was pointed at the region from 17:00 UT to 19:00 UT acquiring filtergrams in the photospheric TiO 7057\AA~  line (bandpass: 10\AA) pixel size of 0.034\arcsec using Broadband Filter Imager (BFI), and narrow band images in the He\textsc{i} 10830\AA~line (bandpass: 0.5\AA) 
with the IRIM instrument \citep{cao2012}. In addition, observations in the H$\alpha$ 6563\AA~line at the blue/red (-0.2\AA/+0.2\AA) wings were acquired with the NST Visible Imaging Spectrometer \citep[VIS,][]{cao2010}. The adaptive optics correction system, AO308, and the speckle image reconstruction processing technique \citep{woger2007} provided diffraction limited images, allowing us to resolve small-scale plasma structures in the photosphere and chromosphere. Time cadence of the processed NST data is 15 s. Simultaneous observations from space telescopes IRIS, SDO and Hinode were also used in the analysis.  The IRIS data included the Mg\textsc{ii}~K line (2796\AA) imaging the upper chromosphere, the S\textsc{iv} transition region (1400\AA), and C\textsc{ii} transition region (1330\AA).  The Atmospheric Imaging Assembly \citep[AIA;][]{lemen2012} and the Helioseismic and Magnetic Imager \citep[HMI;][]{scherrer2012} on board SDO were used for analysing EUV emitting coronal plasma, and the photospheric magnetic field evolution, respectively. G-band and Ca\textsc{ii}~H  filtergrams acquired with the Solar Optical Telescope onboard Hinode were also utilized in the multi-wavelength study, together with the Hinode XRT observations.Data from all telescopes and instruments are carefully co-aligned, and the region of interest spatially and temporally extracted from the different channels.

\section{Results}
The multi-wavelength data allowes us to simultaneously investigate solar events from the photosphere up to the corona. Figure~\ref{figure1} illustrates context of the NST (IRIM and TiO) images (available through on-line Data Catalog at http://www.bbso.njit.edu) displaying a large part of NOAA AR 11810. The figure shows the extracted portions of the chromospheric filtergrams (NST/VIS and IRIS), and coronal (SDO/AIA and Hinode/XRT) images, in a vicinity of a group of solar pores.

\subsection{Chromospheric surge observed in He and H$\alpha$ lines}
\label{ss1}
The NST He\textsc{i} images reveal a very rich chromospheric dynamics and plethora of small-scale activity. An exceptionally enhanced and elongated absorption feature (surge) drew our attention to the region, and we decided to investigate it in detail. Such surges, sometimes more powerful, are quite common \citep{zirin1988}. Figure~\ref{figure1} (upper-leftmost panel) shows the surge (inside the black rectangular box)  at the moment of its maximum intensity contrast. Although the FOV is not large enough to fully cover the length of the surge, its apparent length is at least 23 Mm. Figure~\ref{figure2} is a time-slice plot displaying the evolution of the surge throughout the time series of He\textsc{i} images (various notable evolutionary stages are indicated by a set of arrows). The total lifetime of the surge is approximately one hour. The evolution of the surge shows periods of enlargement and retraction. Possibly, there were three subsequent eruptions, separated by about 30 min. Some dark multi-threaded features are observed, in particular between the first and second eruptions. The limitations in the FOV did not allow us to determine the entire extension of the surge, but the outbursts seem to follow the same trajectory, which may indicate an oscillatory behavior. Up to a dozen of superfine threads can be individually resolved in the surge in some of the images. The maximum thickness of the surge was about $\sim$2.2 Mm, and the width of every resolved fiber is from 100 to 200 km. From the longitudinal growth shown in the plot  we can estimate the longitudinal expansion velocity of about 30 km s$^{-1}$ for the first ejection and $\sim$10 km s$^{-1}$ for the following retraction-ejection periods. The chromospheric response peaks about 3 min after the initiation of the surge. This will be discussed in detail in Section~\ref{ss3}. In the NST/H$\alpha$ images taken in the blue and red wings, the corresponding dark-absorbing feature appears very enhanced (see images in Figure~\ref{figure1}), and the images clearly manifest that the surge runs parallel to other much brighter fibrilar structures in the region, reflecting the geometry of the ambient magnetic field.

\subsection{Photospheric activity and magnetic flux emergence}
\label{ss2}

\begin{figure*}[ht]
  \begin{center}
    \leavevmode
      \includegraphics[width=1.0\linewidth,angle=0,trim = 0mm 0mm 0mm 0mm, clip]{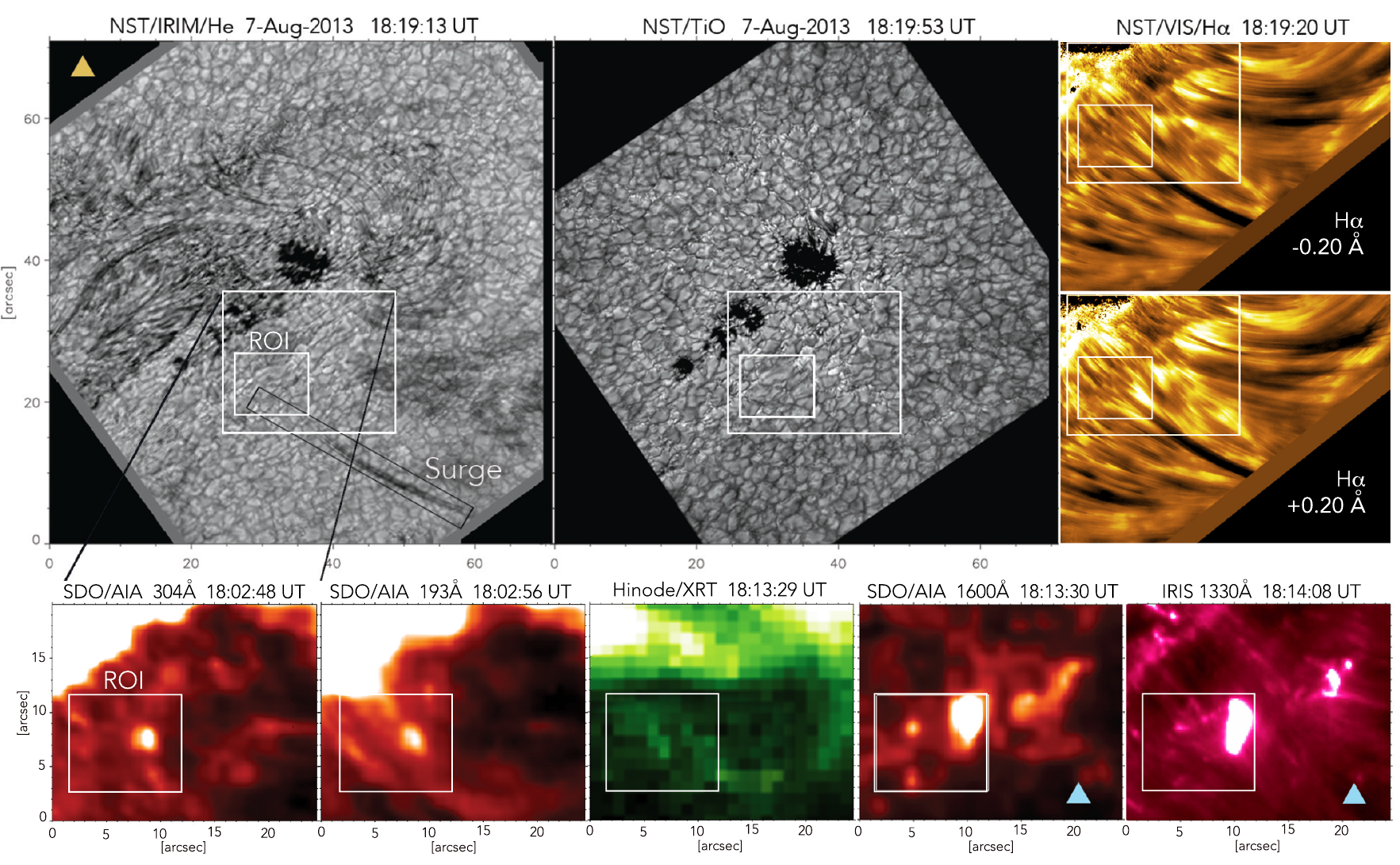}
       \caption{Selected images of NOAA 11810 observed on August 7, 2013, with ground-based (NST) and space telescopes (SDO, IRIS and Hinode).     \emph{Upper row}: NST/IRIM/HeI image (left panel) displaying a dark-absorbing feature (surge) and quasi-simultaneous NST/TiO photospheric image (middle panel), and NST/VIS/H$\alpha$ images in the blue/red wing (right panels). \emph{Lower row}: Chromospheric and coronal observations of the region framed by large white boxes in the upper panels. Images correspond to instances of peak intensity values in the sub-regions framed by the small white boxes. Equal color palette are used to display images observed with the same instrument/telescope. All units are in arcsec.}
     \label{figure1}
  \end{center}
\end{figure*}

\begin{figure*}[ht]
  \begin{center}
    \leavevmode
      \includegraphics[width=1.0\linewidth,angle=0,trim = 0mm 0mm 0mm 0mm, clip]{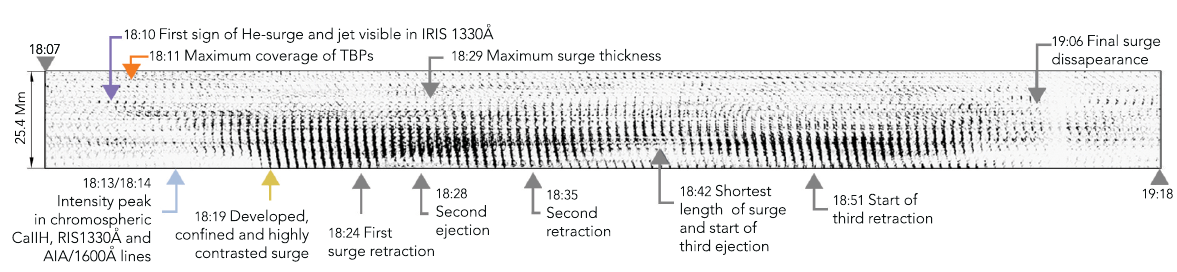}
       \caption{Time-slices plot displaying the evolution of the chromospheric dark-absorbing feature (surge) from the NST/IRIM/HeI observations. Black rectangle framing the surge in Figure~\ref{figure1} (top left panel) shows the portion extracted from the time series to generate the plot with a time interval of 30 seconds between every pair of images. The figure highlights remarkable evolutionary stages throughout the lifetime of the surge, as outcoming from the analysis of all different instruments and channels. Slices with colored arrows denote various stages of the event that are displaying as individual images labeled with the corresponding $\triangle$ in Figures~\ref{figure1}, \ref{figure3} and \ref{figure5} in this paper.}
            \label{figure2}
  \end{center}
\end{figure*}

\begin{figure*}[ht]
  \begin{center}
    \leavevmode
      \includegraphics[width=1.0\linewidth,angle=0,trim = 0mm 0mm 0mm 0mm, clip]{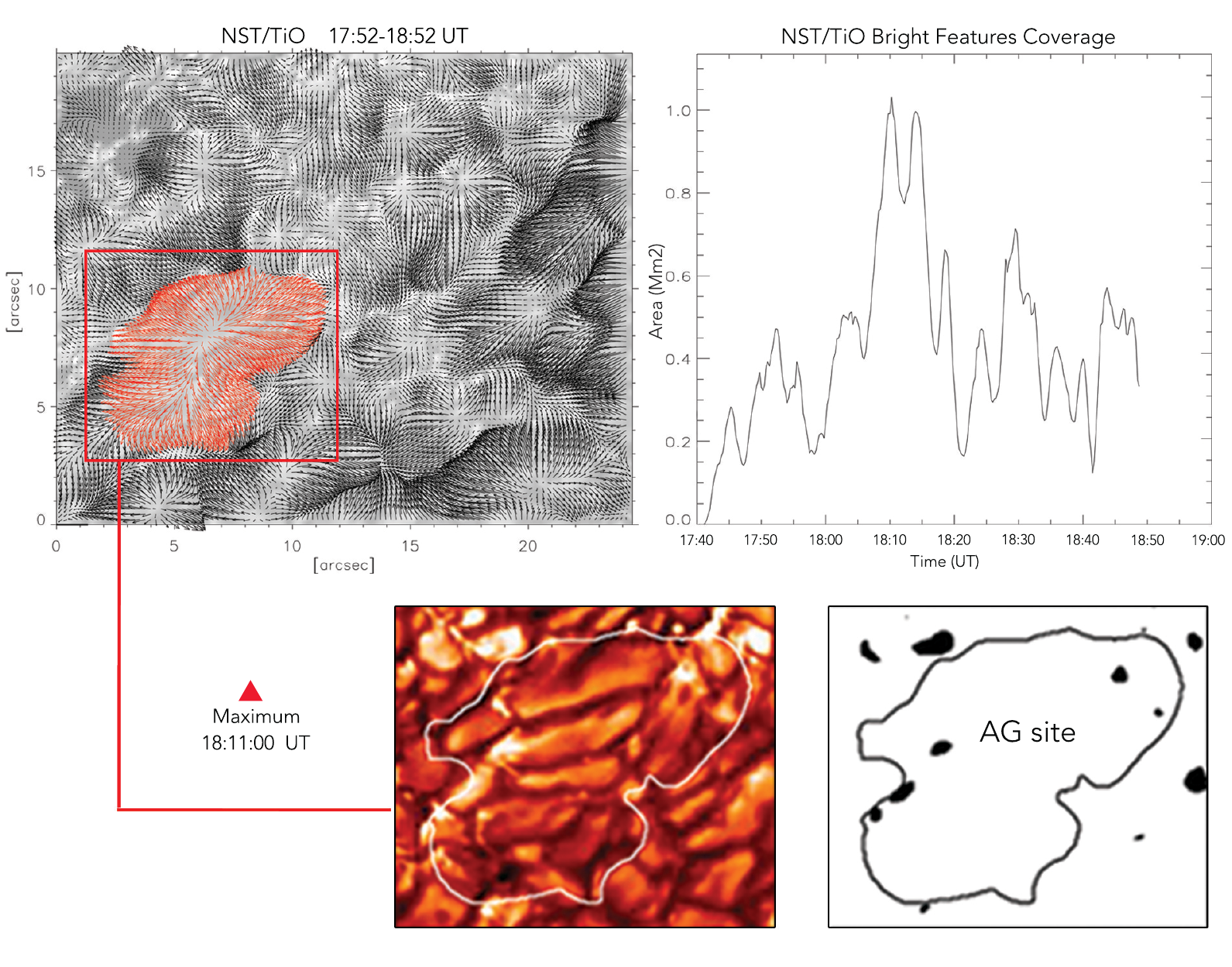}
       \caption{\emph{Upper row}. \emph{Left:} Map of horizontal velocities from the LCT analysis applied to the NST/TiO time series with a 1-hour averaging (17:52 to 18:52 UT). An intense region of divergence (highlighted with red arrows) corresponds to the abnormal granulation (AG) site where exploding and very distorted granules are dominant. Red box extracts a subregion for further analysis in this paper. \emph{Right:} The area covered by bright features in the photospheric TiO images in the red box.  \emph{Lower row}. Panels extract the corresponding image with the maximum number of bright points (\emph{left}) as shown in the mask where black spots are the detected bright features (\emph{right}). }
     \label{figure3}
  \end{center}
\end{figure*}

\begin{figure*}[ht]
  \begin{center}
    \leavevmode
      \includegraphics[width=0.6\linewidth,angle=0,trim = 0mm 0mm 0mm 0mm, clip]{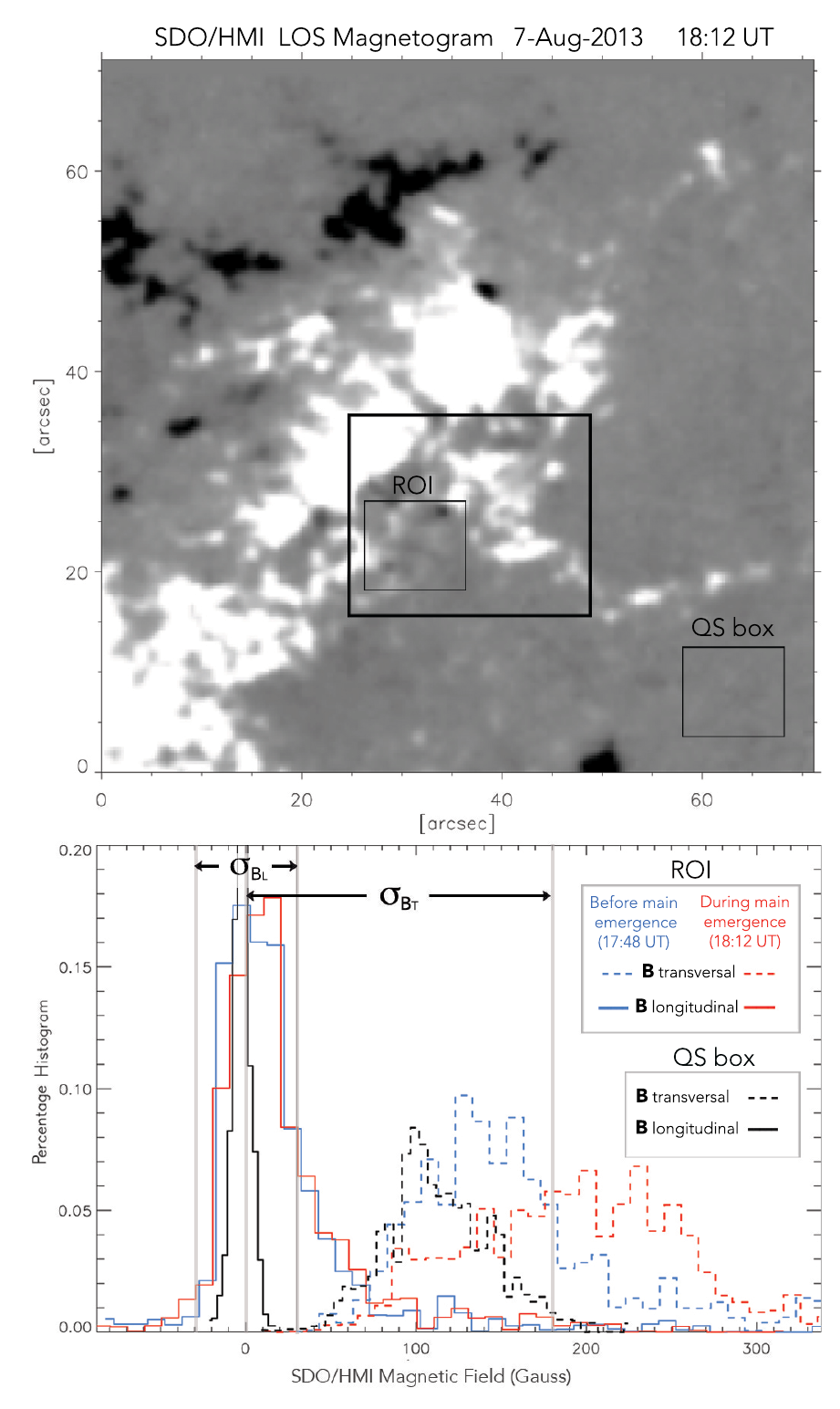}
       \caption{\emph{Upper panel:} LOS Magnetic field from SDO/HMI observations on Aug 7, 2013 at 18:12 UT (about the time of the surge inception, see the text for details). The magnetogram displaying positive/negative magnetic fields (white/red) is saturated to values of $\pm$ 200 G. The FOV is the same as the one in the larger panels in Fig.\ref{figure1}, and so are the extracted boxes in the central part. A region of quiet sun is selected for further comparisons (QS box). \emph{Lower panel:} Histogram of transversal (\emph{dotted lines}) and longitudinal  (\emph{solid lines}) components of the magnetic field computed from the SDO/HMI data. Calculations are done within the ROI before  (\emph{in blue}) and after (\emph{in red}) the emergence of magnetic field and associated distortion of the granular pattern. Values are also calculated for the region in QS box (\emph{black dotted/solid lines}) accordingly. Regions corresponding to noise values in the SDO/HMI data for longitudinal ($\sigma_{B_{L}}$) and transversal ($\sigma_{B_{T}}$) magnetic field are shown.}
       
     \label{figure4}
  \end{center}
\end{figure*}

\begin{figure*}[ht]
  \begin{center}
    \leavevmode
      \includegraphics[width=0.8\linewidth,angle=0,trim = 0mm 0mm 0mm 0mm, clip]{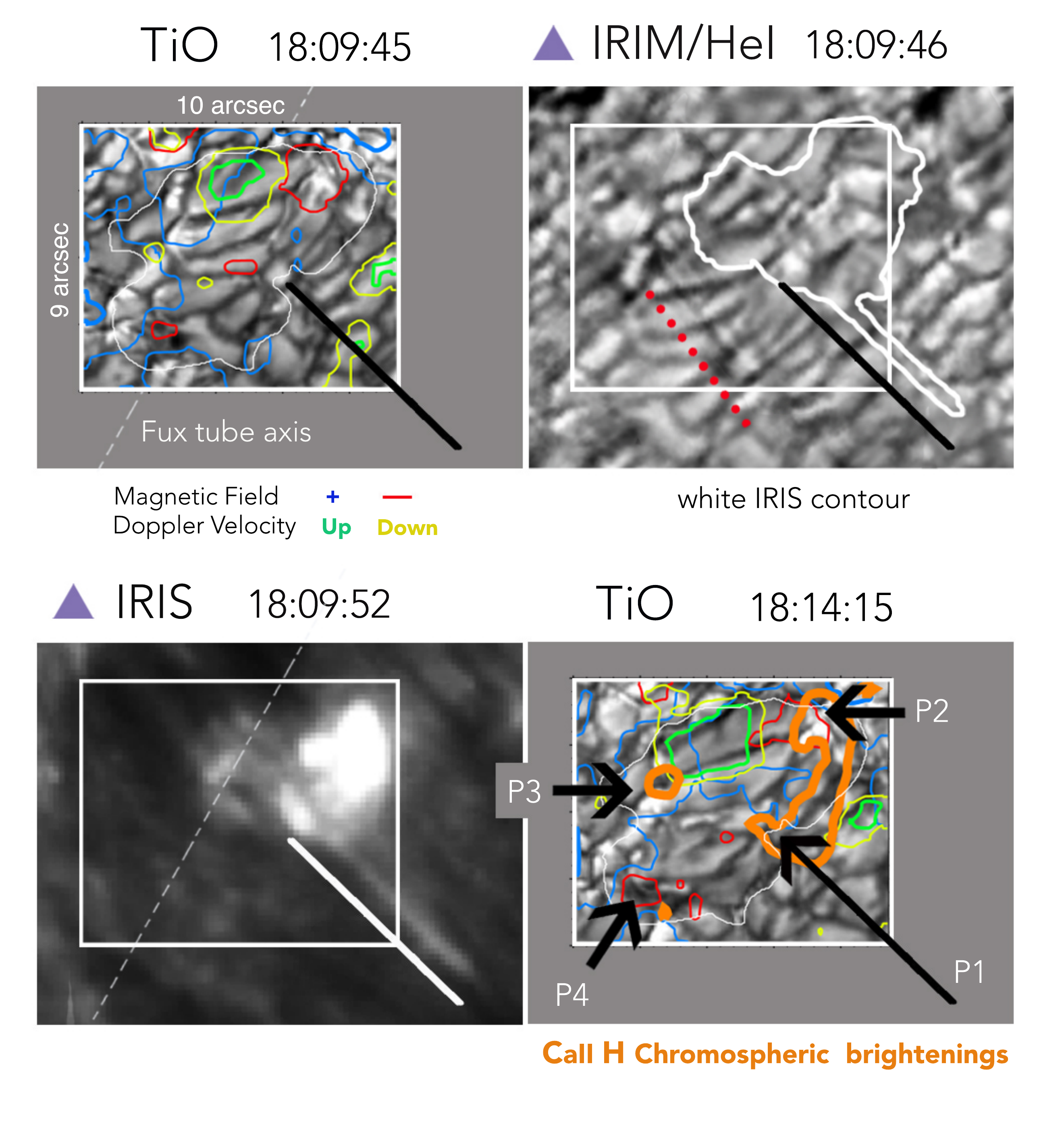}
       \caption{Sequence of selected NST (TiO and He\textsc{i}) and IRIS images displaying the evolution of the region where abnormal granulation is detected (AG site; see text for details). The large FOV corresponds to the small white boxes in the upper panels in Figure~\ref{figure1}, same boxed portion in Figure~\ref{figure2}(left panel). Colored contours are extracted from SDO/HMI and represent the LOS- magnetic field ($\pm$40 G) and Doppler velocities ($\pm$160 m s$^{-1}$) according to the color code shown. White contour in the TiO images represent the AG site. IRIM/He\textsc{i} is shown at the time the jet is visible in IRIS 1330\AA~(right). IRIS contour is overplotted in white in the IRIM frame. In the right frame, arrows indicate the location of points of interest (P1, P2, P3 and P4). Inclined solid lines are shown as references to compared with IRIS image where the jet is observed. The axis of the emerging flux tube has a direction marked by the inclined dashed lines in the first and third frames. The direction of the ambient field can be inferred from the IRIM image (see dotted red line). Purple triangles indicate these images are used to describe the evolutionary plot of the surge in Figure~\ref{figure2}. Time labels are in UT.}
     \label{figure5}
  \end{center}
\end{figure*}

The TiO images show an intense activity of bright features, particularly in the nearest vicinity around the pores (see upper-mid panel in Figure~\ref{figure1}). We find a good agreement in the location of these bright structures (TiO Photospheric Bright Points) and the G-band bright points observed in simultaneous Hinode/G-band images. These bright structures host magnetic flux tubes of the order of kG \citep{ishikawa2007}. The enhanced-brightness photospheric structures match the locations where He\textsc{i} dark-absorbing features seem to be rooted, as pointed out in some previous works \citep{zeng2013,yurchyshyn2010}. At about 18:00 UT, time series of the photospheric TiO images reveal an area of intense Abnormal Granulation (AG) occurring immediately before and during the development of the surge, suggesting a connection between the two phenomena. To track the photospheric plasma dynamics we applied the local correlation tracking method \citep[LCT;][]{november1988} to the series of TiO images. Figure~\ref{figure3} (upper left panel) shows a flow field map computed over the large white-boxed region in Figure~\ref{figure1} (upper panels) by using a correlation tracking window with FWHM of 1\arcsec and averaging over 1-hour time period. The flow field is dominated by a large area of diverging flows (highlighted  with red arrows). Prior to the LCT analysis the time series were subsonic filtered to eliminate p-modes of solar oscillations. Arrows in the figure represent horizontal velocities, and although the velocity magnitude depends on the time averaging, the general pattern can be used to track the surface flows \citep[e.g.][]{vargas2008}. Different 20-min time averaging windows were used to compute the flow maps before and after the interval shown in Figure~\ref{figure3} (17:52--18:52 UT). Only the flow maps calculated during this time interval display the large diverging region. During this time the granule exploded and grew up to five times the size of a normal granule. The region of red arrows is hereafter referred to as the AG site. Some intense photospheric bright points are observed in the closest vicinity, inside and around the AG site. We used an intensity thresholding procedure for masking the location/area of bright features. Figure~\ref{figure3} (upper right panel) plots the variation of the area (Mm$^{2}$) covered by bright features. For further analysis we extracted a Region 
Of Interest (ROI) around the AG site (smaller white box in Figure~\ref{figure1}). Small panels in Figure~\ref{figure3} (lower row) show the extracted TiO frame at the time of the maximum area coverage (18:11:00 UT), and the corresponding mask (where black areas show the location of bright features in the FOV). The black contour outlines the AG site found in the flow map. The time evolution in the plot shows a sharp increase in the area of bright features starting 10 min before the AG activation and up to the maximum (17:42 to 18:11 UT). The time of the area maximum coincides with the very first signs of the plasma surge in He\textsc{i} 10830\AA~line shown in Figure~\ref{figure2}, and is close to the moment of the jet enhancement observed in the IRIS data. In the next section we will comment more on the chromospheric and coronal response. By using the SDO/HMI data we can follow the evolution of the photospheric magnetic field and Doppler velocities in the AG site. Figure~\ref{figure4} shows one of the context LOS magnetograms (saturated at 200 G), in which positive magnetic polarities (in white) are dominant, and correspond mainly to the magnetic field of the pores. Direction of the ambient magnetic fields can be inferred by from the NST chromospheric images (also in IRIS and Hinode images, e.g panels in Figure~\ref{figure1}), and they display organized quasi-parallel spatial distribution. The ROI displays some negative-polarity magnetic elements, see for instance the ROI box in the magnetogram in Figure~\ref{figure4} close to the moment of the detected surge eruption. From the SDO/HMI observations we have vector magnetic field data (magnetic field strength, inclination and azimuth angles)  every 12 min from 17:00 to 19:48 UT.  We calculated the transversal and longitudinal magnetic fields and tracked their evolution. Figure~\ref{figure4} (\emph{bottom panel}) plots histograms for the transversal and LOS magnetic field components computed for two instances: before and during the main emergence of magnetic flux (17:48 and 18:12 UT respectively). Comparing the magnitude of longitudinal magnetic field (above the noise level) we did not find significant changes for both intervals, whereas there is a substantial increase of the transversal component from the first to the second instance (increment of $\sim$100 G in the mean value). We also used a reference box extracted from a quiet Sun region (QS box) to measure the same quantities and found that the distribution of horizontal magnetic field strength is shifted to lower values compared to those for the ROI box (longitudinal components are predominantly below the noise level).
Figure~\ref{figure5} (upper-left and lower-right panels) shows two selected TiO images with overlying contours of positive/negative (blue/red) LOS magnetic field. Contours of Doppler velocities are overplotted in both frames (upflows/downflows in green/yellow). The white contour outlining of the AG site is included as a reference. Regions of weaker magnetic field in the lower part of the FOV display larger upflows and downflows up to 400 and 1400 m/s respectively. The FOV is dominated by positive magnetic fields with maximum mean values of $\sim$300 G throughout the entire time coverage (17:32 to 18:51 UT). More intense photospheric bright points host stronger magnetic field reaching almost kG values. By the moment of initiation of the AG event, inferred from the flow map, confined upflows surrounded by downflows start to appear with bipolar magnetic patches neighboring them (at 17:53:15 UT). During the emergence of the magnetic flux the area of upflows increases together with growing regions displaying positive/negative magnetic polarities. Approximately 15 min after the initiation, the granules become very distorted and ``sandwiched'' forming a lineal structure (see TiO frame at 18:09:45 UT in Figure~\ref{figure5}). By this time, an enhanced jet structure was detected in the IRIS 1330\AA~data together with the first signs of the surge in He\textsc{i} images (Figure~\ref{figure2}), as shown in the corresponding IRIS panel in Figure~\ref{figure5}. The magnetic flux emergence process appears to continue for another 35 min with recurrent upflows with neighbour downflows. Fragmentation of polarities was visible at some stages of this process, but in general, upflows were encompassed by positive and negative small-scale (2\arcsec) patches with the field strength of 50 G. Location of some particular regions in the FOV are labeled as P1, P2, P3 and P4 (see the lower right panel in Figure~\ref{figure5}). P1 indicates the location where the surge eruption initiates (as inferred from the time-slices plot in Figure~\ref{figure2} and the IRIS image in Figure~\ref{figure5}). Points P2 and P4 are characterized by localized negative magnetic patches, while P3 by a more extended positive polarity area. There is a significant increase of the magnetic field strength in P2, P3 and P4 as the region evolves. From 18:20 UT onwards, the negative patch at P4 shrinks and seems to be canceling out with a neighbouring large positive magnetic area while negative path at P3 does the same with its surrounding positive patch. At the end of the sequence (18:51:48 UT) the upflows vanished, and a large patch of positive polarity covered a substantial part of the FOV. The rapid evolution of the magnetic field and plasma vertical velocities occures mostly inside the region described as AG site (within the white contours in the sequence of TiO images in Figure~\ref{figure5}).

\subsection{Chromospheric and coronal response}
\label{ss3}

Simultaneous observations from IRIS, SDO and Hinode allowed us to track the response of the chromospheric/corona to the flux emergence event. Time sequences of images displaying the evolution of the solar atmosphere were generated from 19 spectral different channels. Figure~\ref{figure6} (upper row) shows the intensity variations (light curves) in some of the channels. The SDO/AIA time interval spans from 17:00 to 20:00 UT. The profiles are normalized to their peak intensities. The coronal response is the strongest at 18:03 UT (304 and 193\AA) whereas chromospheric emission measured in 1600 and 1700\AA~ AIA channels, peaked at 18:13 UT. These two moments were used as references for the other panels in Figure~\ref{figure6} (indicated by vertical lines). For the IRIS and Hinode data the peak values were achieved around the same time. We recall the inception of the jet/surge occured at 18:09 UT, between the two reference times. The coronal maximum response occured 6 min before the jet/surge first appearance, and prior to the peak of localized heating of chromospheric layers. The chromospheric Ca\textsc{ii}~H observations, acquired less than 5 min after the jet ejection, reveal the enhanced ultrafine loops of plasma heated at this atmospheric level. The shape of the emerging arch filament system can be traced from the filtergrams (see orange contours in frame 18:14:15 UT in
Figure~\ref{figure5}). Location of P2, P3 and P4 correspond to footpoints where the most prominent emerging loops seem to be rooted, in agreement with
the presence of intense photospheric bright points. The compact coronal brightening is confined to the location of the jet (P1), whereas the chromospheric brightenings cover a curved path connecting the location of the jet eruption (P1) and a negative polarity patch with intense photospheric emission in TiO images (P2). The direction of the jet/surge is parallel to the overlying ambient field lines as marked by 
the \emph{red-dotted line} in the upper right image in Figure~\ref{figure5}. The IRIS image in the same figure shows that the direction of the ejection is almost perpendicular to the axis of the emerging flux tube (\emph{white-dotted line}). At some frames in the Ca\textsc{ii}~H time sequence, before the detected plasma ejection, some dark round-shaped areas (cool patches) are visible with a size resembling the photospheric granular pattern.

\section{Comparison with models}

\begin{figure*}[ht]
  \begin{center}
    \leavevmode
      \includegraphics[width=1.0\linewidth,angle=0,trim = 0mm 0mm 0mm 0mm, clip]{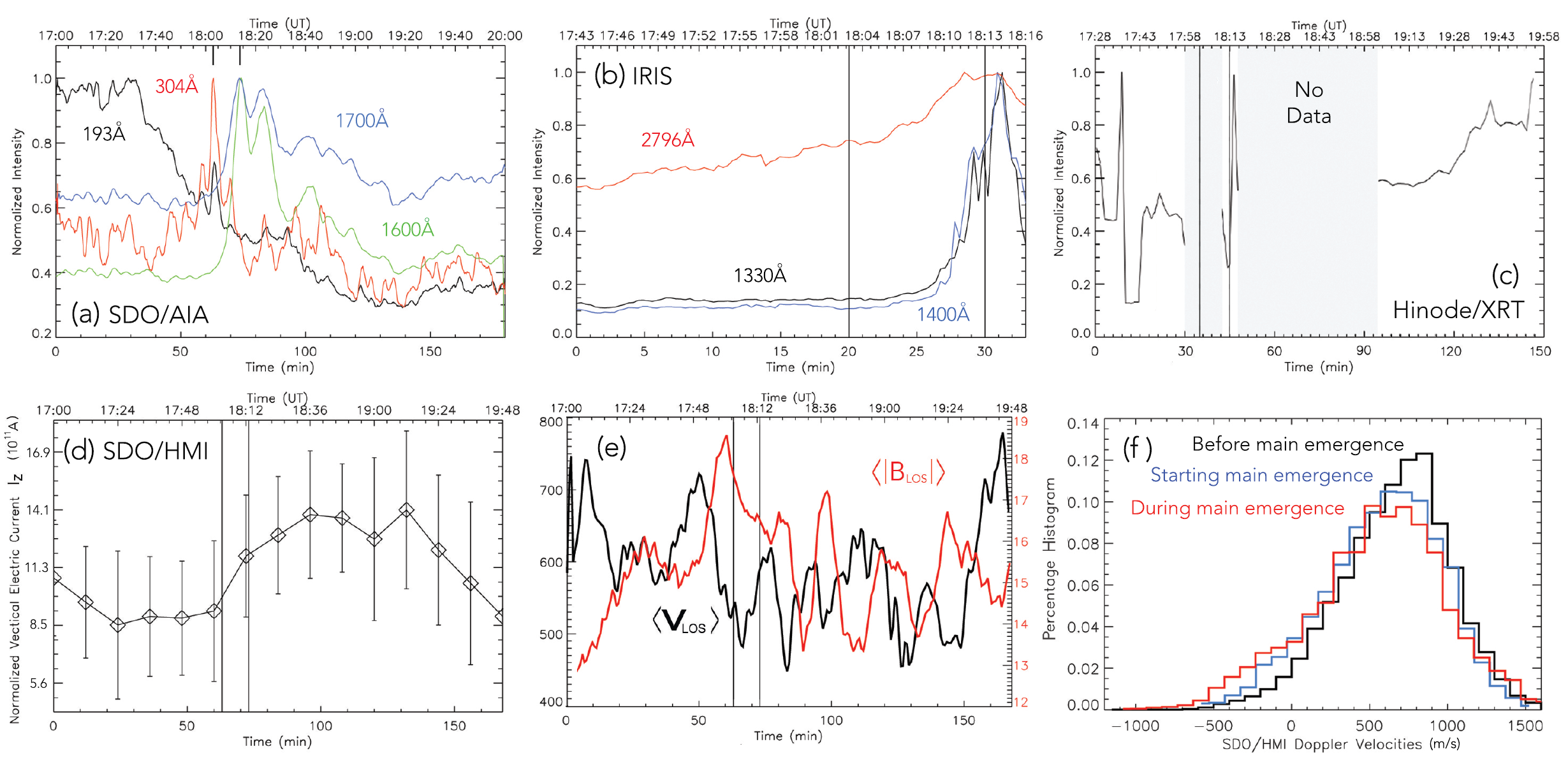}
       \caption{\emph{Upper row}: (a) intensity profiles computed over the time series of SDO/AIA, (b) IRIS and (c) Hinode/XRT. Colored curves indicate different channels as labeled. Values are normalized to the maximum one during every corresponding time period. \emph{Lower row}: (d) variation of mean vertical electric current Iz. (e) evolution of mean LOS Doppler velocity (\emph{black}, [m s$^{-1}$]) and mean unsigned LOS magnetic field (\emph{red}, [G]) from SDO/HMI photospheric data. (f) distribution of vertical velocities before, starting and during the main emergence process. The two vertical lines in panels (b)-(e) are placed at 18:03 and 18:13 UT, as extracted from the intensity peaks in panel (a).}
     \label{figure6}
  \end{center}
\end{figure*}

Magnetic field emergence on the granular and mesogranular scales as well as exploding granules have been 
studied both observationally \citep[][and references therein]{palacios2012} and using
numerical MHD simulations \citep[e.g.][]{cheung2007,cheung2008,martinez2008,stein2012}. Multi-wavelength high-resolution data from several ground and space-based instruments gave us an opportunity to investigate in detail the structure and evolution of these events from the photosphere to the corona, and 
compare the observational results with state-of-the-art simulations. The morphology of the distorted-granulation pattern generally
agrees with emergence of a buoyant horizontal magnetic flux tube from the upper convection zone across
granular convection cells. Differing from the case of an $\Omega$-loop shaped tube with the buoyancy concentrated in 
a localized part in which the distorted granules form a cluster coinciding with the upper part of the $\Omega$ loop, 
rise of a horizontal tube is observed crossing the photospheric level while forming a lane.  Therefore in the last case, large and dark granules appear when the magnetic domain reaches the
photosphere and they are organized along a more longitudinal arrangement that demonstrate the geometry of the initial magnetic tube
\citep{cheung2007}. The 18:09:45 frame  in Figure~\ref{figure5} displays a sequence of distorted and squeezed granules. 
The axis of the emerging magnetic flux tube was almost perpendicular to the direction of the surge as well as to the
orientation of the overlying ambient field lines, as commented in the previous sections. Prior to the development of the anomalous
granulation, weaker and fragmented magnetic patches were detected in that region, in accordance with
the previous initial arrival of the less-strong magnetized components at the more external part (periphery) of the flux tube. Once the
more magnetized volume of a flux tube reaches the photosphere, an excess of magnetic pressure can lead to abnormal granules \citep[][hereafter TM2009]{tortosa2009}.
Our observations support this scenario with abnormal granulation covering an estimated maximum area of $\sim$50 Mm$^2$ (with an area of individual granules up to 8 Mm$^2$, which is very similar to the value of $\sim$7.5 Mm$^2$ obtained by TM2009), and lasting for about 45 min (18:00 to 18:47 UT).  The observed darker and wider intergranular lanes formed by exploding abnormal granules are in agreement with the simulations and also with other observations \citep{otsuji2007}. Downflows are well correlated with the location of intergranular lanes, yet in some scarce cases they are co-spatial with photospheric bright points. In general, the population of vertically magnetized pixels is more numerous in the intergranular lanes, including new lanes that are created in the process of fragmentation of abnormal granules. Downflows are dominant during the entire time series, as reported to be the case in plage regions \citep{ishikawa2008}. Panel (e) in Figure~\ref{figure6} plots the variation of the mean LOS Doppler velocity (\emph{black line}) and the evolution of the mean unsigned LOS magnetic field (\emph{red line}) within the AG site. There is an increase in the unsigned magnetic flux peaking at 18:02 UT, close to the starting time period used to calculate the horizontal flow map in Figure~\ref{figure3}, i.e at a time of strong diverging motions coming from the exploding abnormal granules. A very steep intensification of unsigned magnetic field occurred between 17:39 and 18:01 UT. This seems to be the main emergence responsible for the generation of abnormal granulation, strong diverging plasma flows, and for bringing up the strongest amount of magnetic flux (unsigned) to the photosphere. Prior to this intensification of magnetic flux brought to the photospheric surface the evolution of the mean Doppler velocity shows a sharp rise (beginning at about 17:38 and peaking at 17:51, i.e 10 min before the same behavior is detected in the variation of unsigned magnetic field), as reported 
also in some studies of the photospheric dynamics of emerging magnetic flux \citep[e.g.,][]{kosovichev2009}. After the so-called main emergence phase, i.e. $\sim$2 min after the peak in the unsigned magnetic field, we identified the period of maximum coronal and chromospheric response (see Section~\ref{ss3}), as marked by the two vertical lines in panel (e) in Fig.~\ref{figure6}). The different periods of intensification/decrease of unsigned magnetic flux are possibly related to the different stages in the emergence of the flux tube (TM2009), and cancellation between emerging polarity patches with ambient opposite-polarity areas that we detect during the evolution of the region. To analyze the distribution of vertical velocities we plot histograms over 10-min windows for three periods: before, starting and during the main emergence. Panel (f) in Figure~\ref{figure6} shows the corresponding plots where, in general, the positive velocity values (downflows) are not changing much, in contrast with the negative velocities (upflows) reaching much higher magnitudes during the emerging phase compare to the period before emergence begins (i.e. maximum velocity magnitudes of upflows doubles from about 500 to 1000 m s$^{-1}$; mean values increase about 200 m s$^{-1}$.) Vertical and horizontal velocity values are in general lower (by 50\%) compared to the numerical experiment of TM2008 output. 

Regarding thermal structuring of the lower and mid chromosphere, a wide variety of features described in simulations, in particular cool patches, hot filaments and high-temperature points reported by TM2009 (found at 700 km above the photospheric level in their simulation run), are all found in our data set (IRIS 1330\AA~ and Hinode/Ca\textsc{ii}~H). These structures are described by these authors as consequences of the emergence of magnetic flux tubes and its interaction with the convective granular motions. The situation described with our observations asserts the emerging magnetic field interacts with the (perpendicularly-oriented) overlying ambient field and generates extra response of the chromosphere and coronal, as evidenced by the cool surge, the hot jet and the response in the light curves and filtergrams from transition region and coronal observations around the time of activation of these highly dynamic features. Localized heating at coronal height occurs $\sim$5 min before the first signs of the surge/jet at the ejection site (P1), whereas the response of transition region/chromosphere occurs $\sim$5 min after it. During that period of about 10 min, vertical electric currents, computed from the SDO/HMI transversal and longitudinal photospheric magnetic field components, experienced a substantial increment (panel d in Figure~\ref{figure6}). Ultraviolet plasma flows have previously been reported occurring earlier, up to 10 minutes, than H$\alpha$ surges \citep{jiang2007}, in agreement with our detection.

\section{Conclusions}

Our findings give new observational insights into the process of emergence of small-scale magnetic flux that rises from the subsurface layers, crosses the photospheric level, deforms the granular pattern, and generates energetic and dynamics responses at different heights of the atmosphere. Heating and plasma acceleration in the chromosphere are boosted by emergence of a  small-scale magnetic volume, resembling the case of a horizontal flux tube rising up from the upper convection zone. One possibility of such response is a reconnection scenario, like the ubiquitous small-scale reconnection argued by \cite{shibata2007}, which perhaps can explain the production of a cool surge and hot plasma jet, together with the heating of localized point-like regions up to a million degrees in the corona.
In this scenario, once the newly-emerging magnetic field reconnects with the existing overlying fields, energy is released and thus the reconnection site is heated.  Hot plasma moves upward along the resulting open field lines and may create the high-temperature jet and emission. Another fraction of the plasma goes downward along the close loop created after the reconnection process and could be responsible for the small-scale brightenings, frequently known as micro flaring events.  Closed loops can also generate magnetic tension. Low temperature surges are caused by the upward motion of cold plasma along open magnetic field lines generated by this tensile force. The relative inclination of the emerging and ambient fields can stimulate reconnection process and the energetic outcome. The He\textsc{i} surge exhibits repeated ejections all along the same trajectories, as previously reported for H$\alpha$ surges \citep{schmieder1984}. The NST He\textsc{i} has demonstrated to be of significant relevance for probing, with unprecedented detail, the response and evolution of the solar chromosphere.

Authors acknowledge the BBSO observing and technical team for their contribution and support. The work was partially supported by NASA grants NNX14AB70G, NNX11AO736 and a NJIT grant.


\end{document}